\documentclass [12pt]{article}
\usepackage{amsmath,graphicx,epsfig,setspace,overcite,a4wide}
\title{\bf \large A Theoretical Study of the Electrochemical Gate Effect in a STM-based biomolecular transistor}
\author{  Stefano Corni\\
\noindent  \it INFM-CNR National Research Center on \\ \it nanoStructures and bioSystems at Surfaces (S$^3$), Modena, Italy \\
\noindent \it \small Email: corni.stefano@unimore.it}
\date{}

\begin{document}
\maketitle
\begin{abstract}
ElectroChemical Scanning Tunneling Microscopy (ECSTM) is gaining popularity as a tool to implement proof-of-concept single (bio)molecular transistor. The understanding of such systems requires a discussion of the mechanism of the electrochemical current gating, which is intimately related to the electrostatic potential distribution in the tip-substrate gap where the redox active adsorbate is placed. In this article, we derive a relation that connects the local standard potential of the redox molecule in the tunneling junction with the applied electrode potentials, and we compare it with previously proposed relations. In particular, we show that a linear dependence of the local standard potential on the applied bias does not necessarily imply a monotonous potential drop between the electrodes. In addition, we calculate the electrostatic potential distribution and the parameters entering the derived relation for ECSTM on a redox metalloprotein (Azurin from {\it P. Aeruginosa}), for which experimental results exist. Finally, we give an estimate of the gating efficiency when the ECSTM setup including Azurin is interpreted as a single biomolecular wet transistor, confirming the effectiveness of the electrochemical gating for this system.  
\end{abstract}

\section{Introduction}
\label{intro}
The Holy-Grail of molecular electronics\cite{nitzan03} is the realization of working transistors based on a single molecule. In particular, to realize a field effect single (bio)-molecule based transistor, three different electrodes should surround the molecule: a source, a drain and a gate. Ideally, source and drain should exchange electrons with the molecule (i.e., they should provide the electric current) without affecting the molecular energy levels, while gate should tune the molecular energy level (via the produced electric field) without exchanging electrons with it. This implies that, when the gate is implemented as a third metallic electrode, it must be microscopically close to the molecule and, at the same time, electrically isolated from it.\cite{datta02} Meeting such requirements represents a formidable experimental task. An elegant alternative solution to the gating problem is provided by the electrochemical gate effect,\cite{gate01} conveniently implemented with ElectroChemical Scanning Tunneling Microscopy (ECSTM).\cite{ecstm}    
 
When standard (i.e., non electrochemical) STM is used to study single redox-active (bio)molecule adsorbed on a solid conductive substrate, the resulting setup is close to that of a field effect transistor (the conductive substrate represents the source and the tip acts as the drain) but the gate electrode is lacking. In ECSTM, the STM tip and substrate are immersed in an ionic solution, and their potentials are controlled with a bipotentiostat w.r.t a reference electrode. Thanks to this bipotentiostat, ECSTM allows to control not only the potential bias between tip and substrate, as in {\it in vacuo} STM, but also another parameter:  the difference between the substrate (tip) potential and the reference potential of the solution. In other words, in ECSTM we can modify the electrochemical potentials of tip and substrate with respect to that of a redox molecule in solution, or, vice-versa, we can tune the energy of the molecular redox level w.r.t. the tip and substrate electrochemical potentials. Since the relative position of the redox level controls the molecular conductance, gating is eventually obtained. 
The features of ECSTM has been exploited by different groups to demonstrate electrochemical gating,\cite{tao96,salerno05,haiss03,lindsay05,tao05,albrecht05,ulstrup05,li06} and thus to propose/implement proof-of-concept single molecule transistors.\cite{kuznetsov02c,kuznetsov04} 

Several theoretical works have been devoted to the study of the conduction mechanisms of redox adsorbate in ECSTM\cite{schmickler92,schmickler93,friis98,kuznetsov02b,kuznetsov02,alessandrini06} and, in general, to STM in water solution.\cite{schmickler95,mosyak96,peskin00,galperin02} 
In the present article, the focus is not on the conduction mechanism itself, but on the electrostatic potential induced by the applied electrode potentials in the tip-substrate gap and on its relations with the electrochemical gate effects. A fundamental role in interpreting experiments on {\it standard} STM and molecular wires is played by the electrostatic potential distribution in the tunneling gap. It has been argued that the feature of the $I(V)$ curve dramatically depends on the fraction of tip-substrate potential bias dropping on the molecule.\cite{datta97}  In ECSTM, it is clear that electrostatic potential must play a relevant role too. This was recognized early,\cite{schmickler95a} however, till very recently,\cite{kornyshev06} not much attention has been given to the effects of the ionic atmospheres on the electrostatic potential distribution induced in the tip-substrate gap by the applied electrode potentials. For example, in interpreting experiments on adsorbate conduction in ECSTM, a linear relation between the fraction of the bias potential acting on a redox center and the geometrical position of the redox center in the tunneling gap has been assumed.\cite{albrecht05,ulstrup05} Is this picture correct for ECSTM of adsorbates? One of the aims of the present article is to answer to this question in a specific case that is particularly ambiguous.\\ 
In addition, the feature of the electrostatic potential distribution in ECSTM are strictly related to a quantity which is fundamental when considering ECSTM experiments as realizations of single molecule transistors: the gating efficiency. In this article, this quantity will be estimated as a function of the system parameters. 

The discussion of electrochemical phenomena such as those controlling gating in ECSTM may become muddled due to the use of the term {\it potential} to indicate different quantities. In the discussion that follows we should be careful in distinguishing between these quantities: the electrostatic potential $\phi$ (physical dimension of an energy divided by a charge); the electrochemical potential $\mu$ (physical dimension of an energy) and the electrode potential $E$ or $V$ (physical dimension as $\phi$, but it is related to a difference of electrochemical potentials, divided by $e$, the modulus of the electron charge). 

The rest of the article is organized as follows: in Sect.\ref{sec:1} we review the basics of the gating mechanism in ECSTM, beginning with the ideal case and describing the necessary refinements associated with the microscopic origin of the process; in Sect.\ref{sec:2} we present the continuum model that we use to describe the potential in the ECSTM setup while in Sect.\ref{sec:3} results are presented and discussed. In particular, a simple relation between the standard potential of the redox molecule in the gap and the applied tip/substrate potentials is derived, the electrostatic potential distribution for a protein in the tip-substrate gap\cite{salerno05} is discussed and the efficiency of an ECSTM transistor based on Azurin is presented. Finally, in Sect.\ref{sec:4} some conclusions are drawn.  

\section{Review of the gating mechanism}
\label{sec:1}
In this section, we present a short review of the electrostatic aspects of the gating mechanism. A description of the gating mechanism in the framework of ECSTM electron tunneling mechanisms can be found, e.g., in Ref.\citenum{kuznetsov00}.

We start our discussion by considering an ideal case. Let us suppose that the substrate (i.e., the source in the transistor nomenclature) is grounded. Thus, all the electrochemical potentials $\mu$ are naturally referred to that of the substrate, $\mu_{sub}$. In other words, imposing a variation to the substrate-reference potential difference is equivalent to change the electrochemical potential of the reference electrode ($\mu_{ref}$).\cite{foot1} When such a variation is imposed (i.e., when $\mu_{ref}$ is changed), the characteristic semi-reaction of the reference electrode is drawn out of thermodynamic equilibrium. 
To recover the equilibrium, the reaction advances or retrocedes, releasing or removing ions from the solution. This change in the ionic balance of the solution modifies $\phi_{sol}$, the electrostatic potential of the solution itself, and, in turn, the electrochemical potential of the ions already present in solution. The new equilibrium is settled when the electrostatic potential variation of the solution ($\Delta \phi_{sol}$) is such that to have compensated for the initially imposed electrochemical potential variations: $-e\Delta \phi_{sol}=\Delta\mu_{ref}$, where $e$ is the modulus of the electron charge (see appendix A for a proof).\cite{foot2}    
The molecule feels now the new $\phi_{sol}$. The energy of each molecular electronic level (in particular of the redox one, i.e., that occupied upon reduction) is thus modified by a quantity $-e\Delta \phi_{sol}=\Delta\mu_{ref}$. 
In conclusion, if the potential difference applied between the transistor source (i.e., the ECSTM substrate) and the gate (i.e., the reference electrode) is changed by a given quantity, then, in this ideal picture, the energy of the redox level of the molecule is changed by the same amount (times $-e$), while the source (substrate) and drain (tip) remained constant. This modification of the redox energy level of the molecule w.r.t. the source (substrate) and drain (tip) level is precisely what gives origin to gating.

\begin{figure}
\begin{center}
\includegraphics*[width=0.5\textwidth]{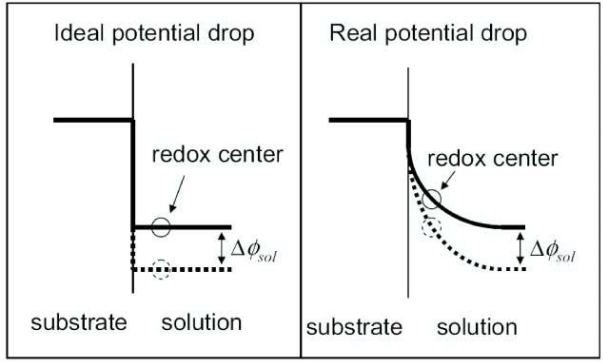}
\end{center}
\caption{Schematic representation of the electrostatic potential drop from the substrate to the solution assuming an ideal (left) or real (right) behavior. Note that when $\phi_{sol}$ is changed, the electrostatic potential at the position of the redox center changes by different amounts in the two cases.}
\label{fig:scheme}
\end{figure}

So far, we did not stress that while $\phi_{sol}$ varies in response to a change in $\mu_{ref}$, the electrostatic potential in the substrate is fixed, since such electrode is grounded. Thus, the electrostatic potential difference between the interior of the substrate and the solution must change of $\Delta \phi_{sol}=\Delta\mu_{ref}/(-e)$. Ideally, there is a step in the potential profile going from the substrate to the solution, whose size is equal to the applied substrate-reference potential up to an unknown but fixed constant. This electrostatic potential drop is step-like only at the macroscopic scale (see Fig. \ref{fig:scheme}). 
In fact, microscopically, the electrostatic potential drop is due to the imbalance of positive and negative ions close to the charged electrode surface, which creates a net density of charge in the solution layer neighboring the electrode.\cite{foot3} 
The structure of this density of charge is described by the Stern double-layer model, which extends for a few nm's (depending mainly on the ionic strength) in the solution. When the redox molecule resides inside this layer (as for adsorbates) the electrostatic potential acting on the molecule (called $\phi_m$) is different from $\phi_{sol}$. Thus, the electrochemical potential variation imposed on the gate (i.e., reference) electrode does not translate completely into the variation of $\phi_m$. A measure the effectiveness of the gate in controlling the molecule potential is the {\it gate control parameter}\cite{datta02}, defined as: 
\begin{eqnarray}
\beta=\left. \frac{\partial \phi_m}{\partial V_g}\right|_{V_{bias}}
\label{eq:beta}
\end{eqnarray}     
where $V_g$ is the potential applied to the gate. Clearly, $\beta=1$ in the ideal case while $\beta \leq 1$ in the real case.

We end this discussion by remarking that the effects related to the incomplete screening of the electrode charge at the position of the molecule have well-known consequences in electrochemical kinetics, where they originate the so-called Frumkin effect.\cite{bockris00,bard}

In the following, we shall discuss how the electrostatic effects of the source and drain electrodes on $\phi_m$ (and thus on $\beta$) can be calculated within a simple model.    
\section{Outline of the model}
\label{sec:2}
We are not aware of calculation of electrostatic potential distribution for a molecule (in particular, a protein) in ECSTM, although calculations for empty tunneling gap appeared very recently.\cite{kornyshev06} Thus, we start by considering a relatively simple model based on a continuum dielectric representation, for which useful relations can be promptly derived. Basically, tip and substrate are represented as perfect conductors (substrate with a planar surface, tip with a spherically curved surface),\cite{foot4} and the adsorbate is a dielectric with a realistic shape (solvent-excluded surface). The redox center of the adsorbate is not explicitly modeled: we simply define $\phi_m$ as the electrostatic potential at the geometrical position of such center. As for the ionic solution surrounding the adsorbate, it is described as another dielectric while the ionic atmosphere is implicitly accounted for by considering the linearized Poisson-Boltzmann equation to obtain the potential in the system:
\begin{eqnarray}
-\nabla\cdot\left( \epsilon \nabla \phi \right)+\epsilon k^2 (\phi-\phi_{sol})= \frac{\rho}{\epsilon_0}
\label{eq:pb}
\end{eqnarray}
where $\epsilon$ is the relative dielectric constant (of the solvent or the adsorbate, depending on the position) and $k$ is the inverse screening length (equal to zero inside the adsorbate), proportional to the square root of the ionic strength $I$. The Helmholtz layer is taken into account by assuming that the region of solution closest to the electrodes cannot contain ions (i.e., $k=0$, see appendix B). The electrostatic potential drop inside this region is thus linear. We shall not consider any specific ion adsorption at the electrode surfaces, i.e., the concentration of ions in the Helmholtz layer is only related to their electrostatic preferences and not to other favorable interactions with the surface. We remark that a reduced dielectric constant for the water layer in the close proximity of the electrodes\cite{bockris00} has been used as discussed in appendix B.

It is clear that the present model is quite simplified and neglects a number of effects, such as the non-linear character of the Poisson-Boltzmann equation, specific ion adsorption and the accounting of the ion finite size in ion-ion interactions. Nevertheless, it catches the main physical features of the ion screening effects (also in complex systems involving biomolecules) and, more important than this, represents the framework in which corrections to the molecular redox level energy as simple as that used in Ref.\citenum{kuznetsov02} can be derived. Refined calculations based on the non-linear Poisson-Boltzmann equation are given and discussed in Sect. \ref{sec:nonl} for a specific system.  

The electrostatic potential that we consider is only due to the applied potential difference: the contributions from the intrinsic charge distribution of the redox adsorbate is already included in the factors that determine the standard redox potential; charging effects of the molecule due to the electronic coupling with the electrodes are not considered here, since we focus on the electrostatic effects due to the charged electrodes. In any case, in the framework of our simple, linear model, these effects are superimposable to our results. 
\section{Results and Discussion}
\label{sec:3}
\subsection{The relation between the electrostatic potential acting on the molecule and the potentials applied to the tip and the substrate}
\label{sec:rel}
In this section, we present a general relation between the electrostatic potential acting on the molecule redox center ($\phi_m$) and the potentials imposed on the electrodes ($E_{tip}$ and $E_{sub}$). Such a relation is based on the continuum model presented above. First of all, we have to specify the electrostatic boundary conditions at the tip and substrate surfaces. We assume that the electrostatic potential at the substrate surface $\phi_{sub}$ is given by $\phi_{sub}=E_{sub}-E_{sub}^{pzc}+\phi_{sol}$, where $E_{sub}$ is the potential imposed to the substrate w.r.t. the reference electrode while $E_{sub}^{pzc}$ is the potential of zero charge (pzc), i.e., the substrate potential at which the surface charge is nil. Analogous definitions are used for the tip, yielding $\phi_{tip}=E_{tip}-E_{tip}^{pzc}+\phi_{sol}$. We remark that the values of $E_{sub}^{pzc}$ and $E_{tip}^{pzc}$ refer to a system where tip and substrate are well separated. Clearly, this does not mean that, e.g., the substrate involved in a STM has a null surface charge at $E_{sub}^{pzc}$, since it is exposed to the electrostatic effects of the other electrode. This is naturally taken into account by the present model. 

Since the equation that defines the potential in our model, eq.(\ref{eq:pb}), is linear in $\phi-\phi_{sol}$, one obtains:
\begin{eqnarray}
\phi_m-\phi_{sol}=\alpha_{tip}(\phi_{tip}-\phi_{sol})+\alpha_{sub}(\phi_{sub}-\phi_{sol})=\alpha_{tip}(E_{tip}-E_{tip}^{pzc})+\alpha_{sub}(E_{sub}-E_{sub}^{pzc}) \label{eq:atas}
\end{eqnarray}   
Thus, the effect of the tip and the substrate potentials on the redox center is, in our simple model, completely described by an intrinsic electrode property, $E^{pzc}$, and by the two coefficients $\alpha_{tip}$ and $\alpha_{sub}$. When there is no electrostatic coupling between tip/substrate and the redox center, $\alpha_{tip}=\alpha_{sub}=0$; when the redox level is pinned to, e.g., the substrate then $\alpha_{sub}=1$. Clearly, $0\leq \alpha_{tip},\alpha_{sub} \leq 1$. In addition, it is possible to proof that $\alpha_{tip}+\alpha_{sub} \leq 1$. Before giving the results of the actual calculations of $\alpha_{sub}$ and $\alpha_{tip}$, we would like to draw a connection with another relation described in the literature to take into account the effects of the tip and substrate potential on the molecular redox levels. To account for such effects in their most recent models,\cite{kuznetsov02,zhang03} Ulstrup, Kuznetsov and coworkers substitute the overpotential $E_{sub}-E^0_m$ (where $E^0_m$ is the standard redox potential for the molecule) with an effective overpotential given by $\xi(E_{sub}-E^0_m)-\gamma V_{bias}$ with $0\leq \xi , \gamma \leq 1$. In ref.\citenum{kuznetsov02}b, $\xi$ and $\gamma$ are defined as "the fractions of the overpotential ($\xi$) and bias voltage ($\gamma$) at the redox site". Following our arguments, in the ECSTM environment the effective redox potential $E^0_{m,loc}$ is given by $E^0_{m}+\phi_m-\phi_{sol}$, thus the effective overpotential becomes $E_{sub}-(E^0_m+\phi_m-\phi_{sol})$. Equating this expression with that used in Ref.\citenum{kuznetsov02} leads to:
\begin{eqnarray}
E_{sub}-(E^0_m+\phi_m-\phi_{sol})=\xi(E_{sub}-E^0_m)-\gamma V_{bias}
\end{eqnarray}
which means:
\begin{eqnarray}
\phi_m-\phi_{sol}=(1-\xi)(E_{sub}-E^0_m)+\gamma V_{bias}
\label{eq:csieta}
\end{eqnarray}  
 In order to compare eq.(\ref{eq:csieta}) with eq.(\ref{eq:atas}), we can recast the latter in the form:
\begin{eqnarray}
\phi_m-\phi_{sol}=\left( \alpha_{tip}+\alpha_{sub} \right) \left(E_{sub}-E^0_m \right)+\alpha_{tip}V_{bias}+\left( \alpha_{tip}+\alpha_{sub} \right) E^0_m -\alpha_{tip}E^{pzc}_{tip}-\alpha_{sub}E^{pzc}_{sub}
\label{eq:csieta1}
\end{eqnarray} 
By comparing eq.(\ref{eq:csieta}) and eq.(\ref{eq:csieta1}), one can identify $\xi=1-(\alpha_{tip}+\alpha_{sub})$ and $\gamma=\alpha_{tip}$. However, there are different terms in eq.(\ref{eq:csieta1}) that does not appear in eq.(\ref{eq:csieta}), playing the role of a correction to the molecular standard redox potential $E^0_m$. Not surprisingly, such a correction depends on $E^{pzc}$ of the tip and the substrate. Notably when $\xi=1$, $\gamma$ must be zero since $\xi=1$ implies $\alpha_{tip}=\alpha_{sub}=0$. 

\begin{figure}
\begin{center}
\includegraphics*[width=0.5\textwidth]{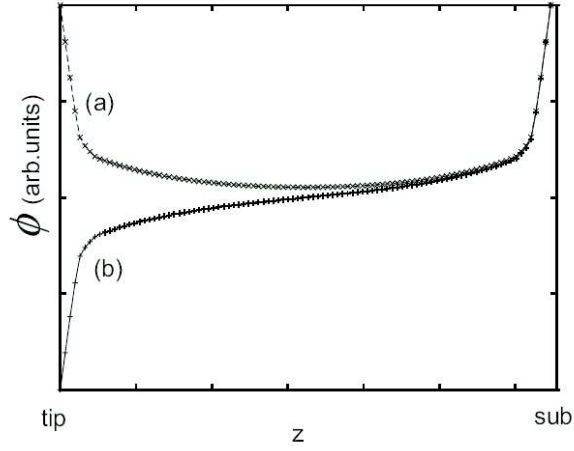}
\end{center}
\caption{Electrostatic potential profiles in pure ionic solution for two different combinations of applied tip and substrate potentials. (a) $E_{sub}>E_{sub}^{pzc}$, $E_{tip}>E_{tip}^{pzc}$; (b) $E_{sub}>E_{sub}^{pzc}$, $E_{tip}<E_{tip}^{pzc}$. Parameters of the calculations: Ionic strength $I=0.05$ M, tip-substrate distance: 40 {\AA}}
\label{fig:non_lin}
\end{figure}

We remark that $\gamma=\alpha_{tip}$ is not directly related to the drop of the applied bias in the tunneling gap, and thus does not reflect the position of the redox center. In fact, $\gamma=\alpha_{tip}$ may be small even when the molecule is closer to the tip than to the substrate, if the ionic concentration is high enough (see below for an example). This is simply understood when one consider the peculiarity of ECSTM w.r.t gas-phase STM. In fact, when STM is performed in gas-phase, the electrostatic potential due to the applied bias is often (and reasonably) assumed to linearly (or at least monotonically) change between the tip and the substrate. In ECSTM, the situation is very different, as demonstrated in Ref.\citenum{kornyshev06}. Suppose that tip and substrate are far enough to allow a complete screening of their respective surface charges by the ionic atmosphere in solution. Depending on these charges (which depend, in turn, on the imposed electrochemical potentials), we can have four different situations, two of which are sketched in Fig.\ref{fig:non_lin} as examples.
It is evident that the existence of a monotonous behavior depends on the applied potentials. What is important to stress here is that even when the potential drop is not linear, a linear relation between $E^0_{m,loc}$ (the standard potential of the molecule in the precise ECSTM gap) and the applied $E_{sub}$ and $V_{bias}$ is obtained from eq.(\ref{eq:csieta1}): 
\begin{equation}
E^0_{m,loc}=E^0_m+\left( \alpha_{tip}+\alpha_{sub} \right) \left(E_{sub}-E^0_m \right)+\alpha_{tip}V_{bias}+E^{corr}
\label{eq:ecorr}
\end{equation}
This means that care must be taken to assume a linear potential drop in the interpretation of ECSTM experiments only on the basis of linearity of the experimental results with the applied potentials. 

In eq.(\ref{eq:ecorr}), $E^{corr}$ is a correction term that does not depend on $E_{sub}$ and $V_{bias}$. It depends on $E^{pzc}_{tip}$ and $E^{pzc}_{sub}$ (whose experimental estimate may be difficult) but it can also collect variations of $E^0_{m,loc}$ that does not originate from the electrostatic effects that we are exploring in this article (due, e.g., to changes in the adsorbate related to the interaction with the electrodes or with other adsorbates or to the different properties of water confined in the nanometric tip-substrate gap).

$E^0_{m,loc}$ is an important quantity in the theories of adsorbate conduction in ECSTM, since it is related to the position of the peak of the current (or apparent height) as a function of the gate potential. Thus, eq.(\ref{eq:ecorr}) is not only useful for theoretical considerations, but also as a tool to interpret the experimental data.  
Remarkably, when the ion screening is ineffective (i.e., $k^{-1}$ much larger than the gap size), one obtains $\alpha_{tip}+\alpha_{sub}\approx 1$ ($\xi=0$) and $E_{sub}-E^0_{m,loc}=-\alpha_{tip}V_{bias}-E^{corr}$ (or $E_{tip}-E^0_{m,loc}=(1-\alpha_{tip})V_{bias}-E^{corr}$), i.e., the relations between the potentials of the electrodes and the effective energy of the molecular electronic levels becomes similar to that used for gas-phase (or non-ionic) STM.\cite{datta97}

\subsection{Electrostatic potential distribution in ECSTM experiments on Azurin}
\label{sec:pot}
\begin{figure}
\begin{center}
\includegraphics*[width=\textwidth]{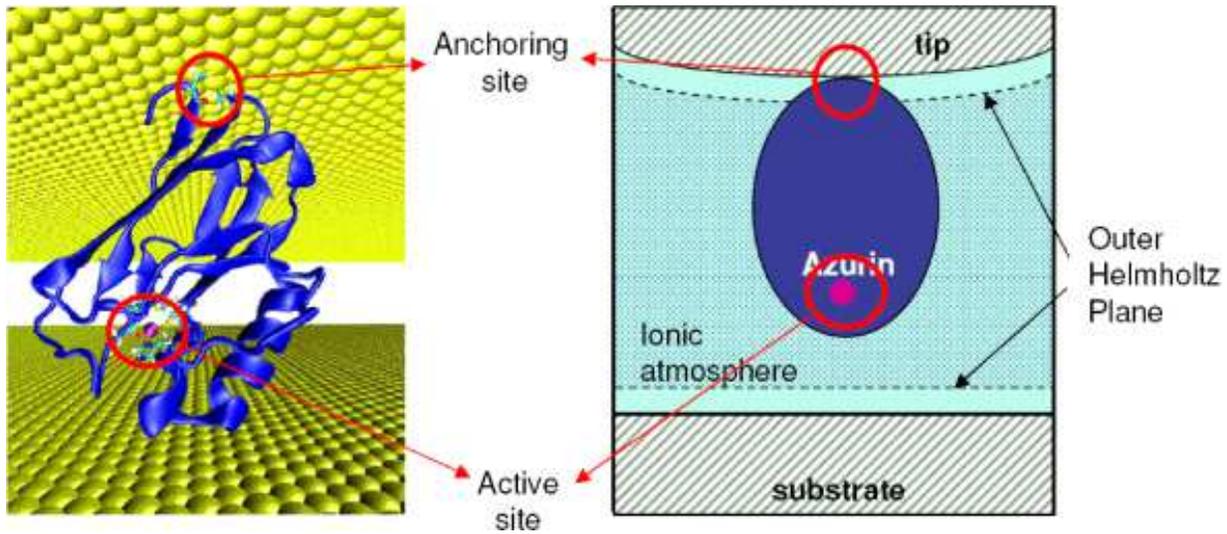}
\end{center}
\caption{Pictorial representation of the Azurin orientation in the ECSTM setup exploited in Ref.\citenum{salerno05}b.}
\label{fig:az_gold}
\end{figure}

\begin{figure}
\begin{center}
\includegraphics*[width=0.5\textwidth]{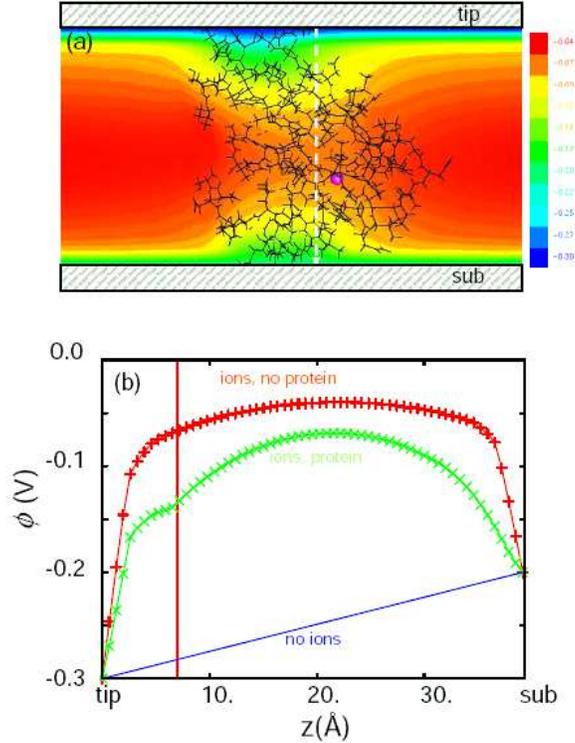}
\end{center}
\caption{Potential distribution for tip-immobilized Azurin in ECSTM. (a) Color-code figures of the potential on a planar surface perpendicular to the electrodes surfaces. (b) Electrostatic potential along the axis perpendicular to the electrodes and crossing the protein (dashed white line in panel (a)). For comparison, the electrostatic potential calculated in ionic solution without the protein ("ions, no protein") and for an homogeneous dielectric filling the space between the electrode ("no ions") are given.  In (b), Azurin occupies the $z$ region on the right of the vertical red line. Parameters of the calculation: $E_{tip}=-0.3$ V$+E_{tip}^{pzc}$, $E_{sub}=-0.2$ V$+E_{sub}^{pzc}$ (thus $V_{bias}=-0.1$ V), ionic strength $I=0.05$ M, tip-substrate distance: 40 {\AA}, tip curvature radius: $\infty$.}
\label{fig:pot_dist}
\end{figure}

\begin{figure}
\begin{center}
\includegraphics*[width=0.5\textwidth]{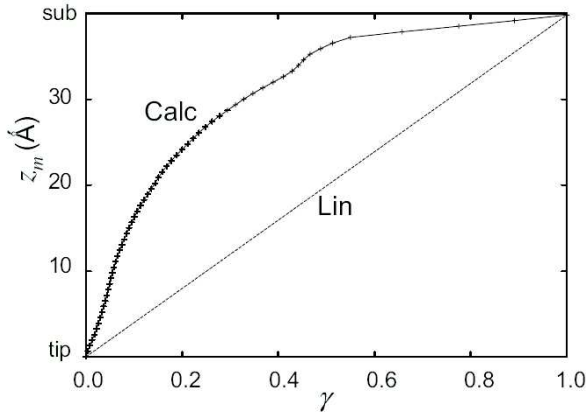}
\end{center}
\caption{Position of the redox center in the tip-substrate gap ($z_m$), estimated on the basis of the $\gamma=\alpha_{tip}$ value. "Calc" refers to our electrostatic calculations, "Lin" refers to a simple linear interpolation. Parameters for the calculation: ionic strength $I=0.05$ M, tip-substrate distance: 40 {\AA}, tip curvature radius: $\infty$.}
\label{fig:pos_comp}
\end{figure}

When the solute is relatively large, like a protein, the electrostatic potential distribution in the tip-substrate gap is more complex than what depicted in Fig.\ref{fig:non_lin}. In fact, the space occupied by the solute is not accessible to the ions, which cannot thus screen the potential due to the surface charges of the tip and the substrate. Thus, the potential drop inside the protein will be intermediate between two limit cases: (a) equal to that in the pure ionic solution; (b) similar to that in a non-ionic dielectric, i.e., linear as in gas-phase. One may thus wonder whether in real experiments, the potential drop is more similar to that in gas-phase or that in ionic solution without the adsorbate. To address this question, we have calculated the potential distribution with the numerical model described in appendix B. In particular, we choose to focus on the recent experiments by Facci and coworkers.\cite{salerno05}$^b$ They studied via ECSTM the transport properties of tip-immobilized Azurin, a blue-copper protein, and they demonstrate a transistor like behavior for the system. In this experiment, Azurin was anchored to the gold tip by exploiting the accessible disulfide bridge between Cys3-Cys26. In our calculation, Azurin has been oriented with the major axis of inertia perpendicular to the substrate, with the S atoms of Cys3 and Cys26 at bonding distance from the tip surface. The protein redox active site (which comprises a redox active copper ion) is on the opposite side of the protein. A pictorial representation is given in Fig.\ref{fig:az_gold}.
This orientation is compatible with the experimental morphology data of Alessandrini et al.\cite{salerno05}$^b$ In Fig.\ref{fig:pot_dist} we report the potential distribution along the axis perpendicular to the electrodes and crossing the protein. To give an idea of the complexity and inhomogeneity of the potential distribution we also report, in the same figure, the potential on a slice in the protein region.  
This figure confirms what we anticipated above: the potential distribution in the protein region is intermediate between the gas-phase and the pure ionic solution cases, and it is closer, in the present case, to that of a pure ionic solution. In particular, we numerically show that the STM-like picture (linear potential drop between the tip and the substrate) is inappropriate to describe the potential drop in ECSTM even for a large ion-excluding adsorbate. These results also suggest that the correlation between the geometrical position of the redox center and the potential acting on it is not straightforward. To better circumstantiate this point, we have taken the hypothetical point of view of someone who measured the $\gamma=\alpha_{tip}$ coefficient from an experiment (e.g., fitting the current as a function of the potentials $E_{sub}$ and $V_{bias}$), and who wants to use this quantity to estimate the geometrical position of the redox center in the tunneling gap (named $z_m$). If one assumes that $V_{bias}$ drops linearly, the position $z_{m}$ will simply be given by $z_m=\gamma\cdot l_{gap}$, where $l_{gap}$ is the gap length (i.e., the tip-substrate distance). We can compare this prediction with what can be obtained by our electrostatic calculations (these calculations give $\gamma(z_{m})$, which can be inverted to yield $z_m(\gamma)$). The comparison is performed in Fig. \ref{fig:pos_comp}.
As it can be seen, the differences can be substantial, and the redox site can be misplaced by the linear interpolation up to 16 {\AA}, which is the 40\% of the tip-substrate gap size. It is also evident that the linear interpolation results always place the redox center closer to the tip than what it really is.  

\subsection{Effects of non-linearity in the Poisson-Boltzmann equation on the potential distribution}
\label{sec:nonl}
\begin{figure}
\begin{center}
\includegraphics*[width=0.5\textwidth]{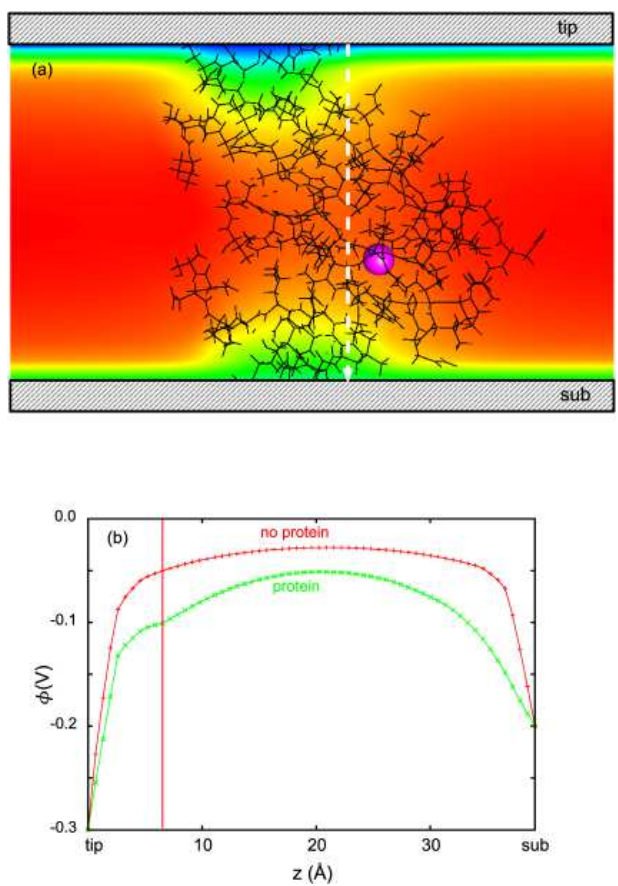}
\end{center}
\caption{Potential distribution for tip-immobilized Azurin in ECSTM calculated from the non-linear Poisson-Boltzmann equation. (a) Color-code figures of the potential on a planar surface perpendicular to the electrodes surfaces (the color scale is the same as in Fig.\ref{fig:pot_dist}. (b) Electrostatic potential along the axis perpendicular to the electrodes and crossing the protein (dashed white line in panel (a)). The electrostatic potential calculated in ionic solution without the protein ("no protein") is also given.}
\label{fig:pot_dist_nol}
\end{figure}

\begin{figure}
\begin{center}
\includegraphics*[width=0.5\textwidth]{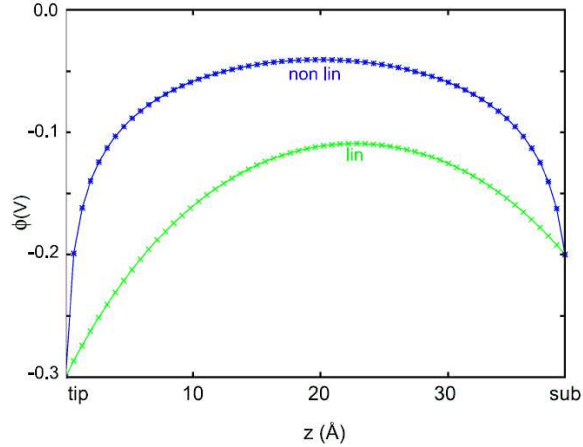}
\end{center}
\caption{Electrostatic potential profile in the tunneling gap calculated in ionic solution without the protein and neglecting the Helmholtz layer effects. "lin" refers  to the potential obtained from the linearized PB equation while "non-lin" refers to the non-linear PB results.}
\label{fig:pot_dist_nohel}
\end{figure}

As mentioned above, the relations and the results presented in Sects. \ref{sec:rel} and \ref{sec:pot} relay, among other assumptions, on the linearization of the Poisson-Boltzmann equation. Due to the potential range (0.2-0.3V) and ionic strength (0.05M) considered, this assumption needs to be check. In the next section, we will show that experimental results for Azurin are indeed linear in the applied potentials, indicating that non-linear effects in the real system should be small. 
Nevertheless, as a consistency check of the model, we have performed some calculations with the non-linear Poisson-Boltzmann equation to give an estimate of the importance of non-linear effects for the studied system. 
First, let us note that in principle, when the assumption of linearity is removed, we cannot consider separately the potential due to the intrinsic density of charge of Azurin and that due to the charged electrode surfaces. 
However, the total charge of Azurin at neutral pH is very small (0 or -1, depending on the oxidation state of Cu),\cite{francy00} and electrostatic potential calculations shows that the molecule produces a potential in solution smaller than $k_BT$, that is well described by the linearized PB equation. Thus, to give an estimate of the effects of non-linearity we have focused on the potential generated by the charged electrodes alone. In particular, we have repeated the 3D calculations leading to Fig.\ref{fig:pot_dist} by exploiting the non-linear PB equation. 
Results are presented in Fig.\ref{fig:pot_dist_nol}.   
As it can be seen, the general appearence of the two figures is very similar, pointing to the fact that the linearized Poisson-Boltzmann equation correctly catches the main features of the potential distribution. The color-coded potential distributions of the two panels (a) can be hardly distinguished.  Note in particular that the relative behavior of the protein vs the ionic solution potential drop is the same in the two figures. In panels (b), the most evident difference is in the region between 0 and $~$7 {\AA}, outside the protein.  On the absolute potential scales, a difference of 15-20 mV ($~5-10$\% of the applied potentials) between the linear and the non-linear calculations is obtained on the the precise values of the potential plateau reached in the middle of the gap. 
The overall agreement between Fig.\ref{fig:pot_dist} and Fig.\ref{fig:pot_dist_nol} can be explained by noting that most of the potential drop takes place in the Helmholtz layers that, being free from ions, behaves similarly in the linear and non-linear cases. Such an idea is supported by calculations that we performed for the empty gap by neglecting the existence of the Helmholtz layers. The results of such calculations are reported in Fig.\ref{fig:pot_dist_nohel}, where the linearized and the non-linearized PB potential drops are plotted together. 
As it can be seen, linearization is a much worse assumption in this case, leading to larger discrepancies between the potential profiles.

\subsection{Comparison with experimental results}
Comparing the theoretical predictions reported in this article with existing experimental results is not an easy task. In fact, the experimental $i(E_{sub},V_{bias})$ curve depends on other parameters beside $\alpha_{sub}$, $\alpha_{tip}$ and $E^{corr}$ that appear in eq.(\ref{eq:ecorr}), such as the reorganization energy and the transfer integrals of the electrode-molecule electron transfer reaction. In principle, to extract all these quantities one can assume a theoretical relation $i(E_{sub},V_{bias})$ and then adjust the parameters entering such a relation ($\alpha_{sub}$, $\alpha_{tip}$, etc.) to reproduce the experimental results. 
However, such a multidimensional fit is not straightforward (different sets of parameters can give similar fitting quality). Moreover, different mechanisms have been proposed to explain the conduction of redox-active molecules,\cite{kuznetsov02b,kuznetsov02,schmickler93} yielding different relations and thus different fitted parameters. Nevertheless, we have exploited the experimental data of Alessandrini et al.\cite{salerno05}$^b$ to extract one quantity that can be compared with our theoretical calculations. We have chosen this experiment among the few available ones for a number of reasons: 
(a) the ionic strength is not too high (0.05M), which improves the quality of some of the model approximations (e.g., linearization of Poisson-Boltzmann equation);
(b) it is one of the few single-molecule ECSTM experiments in which the current as a function of the substrate potential is collected for several different bias potentials; 
(c) it is the system where the non-linear nature of the potential bias drop is less obvious, since the ion screening is ineffective inside the protein and the ionic strength is relatively low; 
(d) the electronic couplings of the electrodes with the redox center is small, thus justifying the neglecting of electrostatic potential changes due to the effects of the flowing current on the molecular density of charge;
(e) the configuration of the molecule with respect to the electrode is relatively well characterized, since azurin has only one chemisorption site available for Au (the Cys3-Cys26 solvent exposed disulphide bridge), and it is a quite rigid protein; 
(f) the size of the tip-substrate gap is, very likely, comparable to the molecular size ($~4$ nm) and thus relatively large. Hence, ions concentration in the gap should not be affected by steric exclusion phenomena.\cite{kornyshev06}

The results of Ref. \citenum{salerno05}b are curves $i(E_{tip})$ at different values of $V_{bias}$. Each $i(E_{tip})$ curve has a maximum for a given value of $E_{tip}$, called $E_{tip}^{max}$. Correspondingly, we can define $E_{sub}^{max}$, which is simply given by $E_{tip}^{max}-V_{bias}$, and $E_{avg}^{max}$ as $(E_{tip}^{max}+E_{sub}^{max})/2$. $E_{avg}^{max}$ represents the mid-point between the potential of the tip and that of the substrate and, in different theories proposed to explain the conduction of adsorbate, it is directly related to $E^0_{m,loc}$ (see Ref.\citenum{alessandrini06} and references therein). In addition, it does not privilege any of the two electrodes, and we shall use it for our theory-experiment comparison. Extracting the value of $E_{tip}^{max}$ per each value of $V_{bias}$ in the experiment, we end up with an experimental $E_{avg}^{max}$ vs $V_{bias}$ curve. On the other hand, from eq.(\ref{eq:ecorr}) it is possible to derive a {\it theoretical} relation $E_{avg}^{max}(V_{bias})$ that can be used to fit the experimental $E_{avg}^{max}$ vs $V_{bias}$ trend by using $\alpha_{tip}$ and $\alpha_{sub}$ as adjustable parameters. The quality of the fit will be a measure of the validity of eq.(\ref{eq:ecorr}), and thus of the linearization of the Poisson-Boltzmann equation, while the fitted values of $\alpha_{sub}$ and $\alpha_{tip}$ will be compared with the estimates based on our numerical model. 
To derive such a theoretical equation, we need a relation between $E^0_{m,loc}$ and $E_{avg}^{max}$. The sequential two-step model (the most likely mechanism for the experiments in Ref. \citenum{salerno05}b, see also \citenum{alessandrini06}) predicts $E^0_{m,loc}\approx E_{avg}^{max}$.\cite{alessandrini06} Introducing such a relation in eq.(\ref{eq:ecorr}) leads to the $E_{avg}^{max}(V_{bias})$ expression that we were looking for:
\begin{equation}
E_{avg}^{max}=\frac{1}{2}\frac{\alpha_{tip}-\alpha_{sub}}{1-\alpha_{tip}-\alpha_{sub}}V_{bias}+E^0_m+\frac{E^{corr}}{1-\alpha_{tip}-\alpha_{sub}}
\label{eq:eloc}
\end{equation}
Fitting the data of Ref.\citenum{salerno05}b, we obtained that the experimental relation between $E_{avg}^{max}$ and $V_{bias}$ is indeed linear as predicted by eq.(\ref{eq:eloc}) (regression coefficient R=0.95) and that the slope $(\alpha_{tip}-\alpha_{sub})/2(1-\alpha_{tip}-\alpha_{sub})$ is -0.12. From our electrostatic calculations, we found $\alpha_{sub}=0.26$ and $\alpha_{tip}=0.13$, yielding $(\alpha_{tip}-\alpha_{sub})/2(1-\alpha_{tip}-\alpha_{sub})=-0.11$, in good agreement with the experiment. Unfortunately, the intercept of eq.(\ref{eq:eloc}) depends on $E^{corr}$, which cannot be directly derived by our simple numerical model.

Clearly, this theory-experiment comparison is quite limited and involves some assumptions (such as the choice of the conduction mechanism). However, we can at least say that available experiments seem not to contradict our predictions. We hope that our results may stimulate further experiments in this field.  

Other experiments on Azurin,\cite{ulstrup05} which employed different ECSTM geometries, cannot be treated with the same procedure used for the data of Ref. \citenum{salerno05} since they do not measure currents but only changes in apparent heights. In addition, they do not explore the $V_{bias}$ dependence of the results.

\subsection{Quantifying the gating efficiency $\beta$ for the Azurin-based ECSTM transistor} 
We now pass to the discussion of the gating efficiency for Azurin in ECSTM. We have already mentioned the gate control parameter $\beta$, which is a synthetic way to express the gating efficiency in a molecular Field Effect Transistor (FET)-like geometry. When $\beta=1$, all the potential applied to the gate is felt by the molecule; vice-versa when $\beta=0$ the gate is not able to modify the redox energy level. When ECSTM is seen as a FET, we can identify the potential of the gate $V_g$ in eq.(\ref{eq:beta}) with that of the reference electrode, measured w.r.t the grounded electrode (the source), meaning that $V_g=(\mu_{ref}-\mu_{sub})/(-e)$. Since $E_{sub}$ is defined above as the potential difference between the substrate and the reference electrode, and the substrate electrochemical potential is fixed (we recall that the substrate is grounded), we can write $E_{sub}=-V_{g}$. Thus, $E_{tip}=E_{sub}+V_{bias}=-V_g+V_{bias}$. In addition, we have noted above that $-e\Delta \phi_{sol}=\Delta \mu_{ref}$, which means that the potential of the solution $\phi_{sol}$ differs from $V_g$ by a fixed but unknown constant $c$ (i.e., $\phi_{sol}=V_g+c$). Substituting these relations in eq.(\ref{eq:atas}), one obtains:
\begin{eqnarray}
\phi_m=c+V_g+\alpha_{tip}(-V_g+V_{bias}-E_{tip}^{pzc})+\alpha_{sub}(-V_g-E_{sub}^{pzc}) \label{eq:vm}
\end{eqnarray}
$\beta$ can be easily derived from eq.(\ref{eq:vm}):
\begin{eqnarray}
\beta=\left. \frac{\partial \phi_m}{\partial V_g}\right|_{V_{bias}}=1-(\alpha_{tip}+\alpha_{sub})
\label{eq:beta1}
\end{eqnarray}
The meaning of eq.(\ref{eq:beta1}) is clear: when the redox center does not feel the surface charge of tip and substrate (i.e., $\alpha_{tip}=\alpha_{sub}=0$), the gating works ideally ($\beta=1$); on the contrary, if the molecule is affected by the tip and/or the substrate charge the gating efficiency is reduced ($0<\beta < 1$). Remarkably, the value of $\beta$ also affects the position of the peak in the tunneling current as a function of the gate potential. In fact, the peak position is related to the local standard potential of the adsorbate, $E^0_{m,loc}$, which depends on $\beta$ via $\alpha_{tip}$ and $\alpha_{sub}$ as illustrated by eq.(\ref{eq:ecorr}). For example, in the sequential two-step model the substrate potential giving the current maximum, $E^{max}_{sub}$, satisfies $E^{max}_{sub}+V_{bias}/2 \approx E^0_{m,loc}$.\cite{alessandrini06}  Introducing such condition in eq.(\ref{eq:ecorr}) one obtains: 
\begin{eqnarray}
E^{max}_{sub}=E^0_m+\frac{\alpha_{tip}-1/2}{\beta}V_{bias}+\frac{E^{corr}}{\beta}
\end{eqnarray}
When $\beta=1$ (i.e., $\alpha_{tip}=\alpha_{sub}=0$), then $E^{max}_{sub}=E^0_m+E^{corr}-V_{bias}/2$, while $\beta<1$ can either increase or decrease $E^{max}_{sub}$ depending of the sign of $E^{corr}$ and the value of $\alpha_{tip}$.  

\begin{figure}
\begin{center}
\includegraphics*[width=0.5\textwidth]{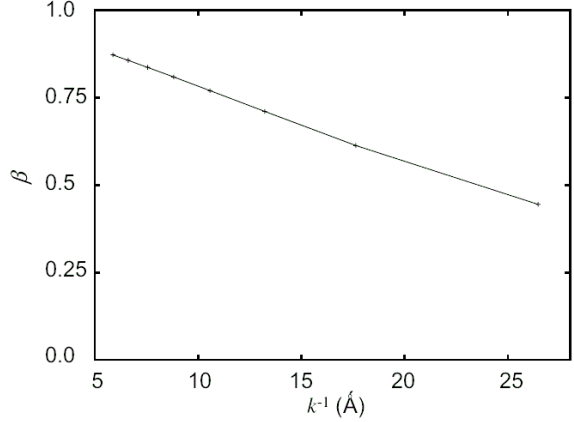}
\end{center}
\caption{The gating efficiency $\beta$ for an Azurin-based transistor as a function of the screening length $k^{-1}$. The tip-substrate distance is 40 {\AA}}
\label{fig:beta_I}
\end{figure}

Since an effective screening of the electrode surface charges is needed to have high values of $\beta$, one expects that an important parameters for determining $\beta$ is the ionic strength $I$ of the solution. In fact, the higher the ionic strength, the smaller is the screening length $k^{-1}$. To estimate (at least in the framework of our simple, linearized model) the dependence of $\beta$ on $k^{-1}$, we have performed calculations on the Azurin-based biotransistor, whose results are depicted in Fig.\ref{fig:beta_I}.
Although the values of $\beta$ indicate that the azurin-based transistor is not working ideally, they are high enough to confirm that the ECSTM approach to gating is convincing even in the ambiguous situation of a protein. In particular, the values of $\beta$ range from 0.45 to 0.85 for $k^{-1}$ between 25 {\AA} and 5{\AA}. The value of $\beta$ for the ionic strength used in the experiment of Ref. \citenum{salerno05}b (I=0.05 M) is around 0.6, which is coherent with the observed gating behavior. In the light of the results of Fig. \ref{fig:beta_I}, we can define $k^{-1}$, the typical length over which the effects of a charge in ionic solution are screened, as an {\it effective gate size} in ECSTM. In fact, $\beta\approx 1$ when $k^{-1}<<l_{gap}$ and $\beta\approx 0$ when $k^{-1}>>l_{gap}$, as expected. Remarkably,  we also found that $\beta=0.5$ when $k^{-1}\approx 20 \mathrm{\AA} \approx l_{gap}/2$ (the tip-substrate distance $l_{gap}$ is 40 {\AA}, which is in turn related to the protein size).

\section{Conclusion}
\label{sec:4}
To summarize the contents of this article, we have discussed how the gating mechanism acts in a wet transistor based on ECSTM and we have used a simple computational model to give numerical estimates of the relevant quantities for a bio-transistor based on Azurin. In particular:
\begin{itemize}
\item We argued that the potential drop (due to the applied bias) in ECSTM of large adsorbates is not monotonic and linear, in analogy to what happens for empty ECSTM gaps.\cite{kornyshev06}
\item We showed that the potentials applied to tip and substrate can indeed affect the redox energy level of the probed biomolecule, when realistic parameters for the system are used.
\item We demonstrated that, in the simplifying framework of a linearized Poisson-Boltzmann model, the variation of the redox energy level is indeed proportional to the tip and substrate potentials, as exploited in models to explain ECSTM.\cite{kuznetsov02} We have emphasized that this proportionality does not generally imply a spatially linear potential drop between tip an substrate, an assumption that may lead to misinterpretations of the experimental results. In addition, we have considered the role of pzc potentials in a straightforward way.       
\item We have quantified the gating control parameter $\beta$ for the protein bio-transistor, an important quantity that expresses the gating efficiency as a function of the parameters of the system. 
\item We have identified a quantity (the screening length $k^{-1}$) that acts as an effective gate size in ECSTM.
\end{itemize} 
We remark again that the computational model is quite rough, and to obtain more precise estimates, more complex models must be used.  However, since this is, to the best of our knowledge, the first study on the potentials in these bio-ECSTM systems, the simple model used allowed a clear indication of trends and gave insights into the basic physics of the system. 

{\bf Acknowledgments} 
The author thanks Andrea Alessandrini and Paolo Facci for the useful discussions and the reading of a previous version of the manuscript.

\appendix
\section{Proof that $-e\Delta \phi_{sol}=\Delta \mu_{ref}$}
Let us suppose that the half-reaction of the reference electrode is in the general form:
\begin{equation}
\sum_j \nu_j J^{c_{j}} - n e^-_{elec} = 0
\label{eq:reac}
\end{equation}
where we have used the convention that stoichiometric coefficient $\nu_{j}$ are negative for reactant and positive for product, while $c_{j}$ is the charge of the species $J$ and $n$ is the number of exchanged electrons. Charge balance of eq.(\ref{eq:reac}) implies:
\begin{equation}
\sum_j \nu_j c_{j} +n = 0
\label{eq:bal}
\end{equation}
while thermodynamic equilibrium is expressed by:
\begin{equation}
\sum_j \nu_j \mu_{j} - n \mu_{ref} = 0
\label{eq:ther}
\end{equation}
where $\mu_{j}$ is the electrochemical potential of the species $J$ and $\mu_{ref}$ is the electrochemical potential of the electrons in the reference electrode. When $\mu_{ref}$ is changed to $\mu_{ref}'$, with $\Delta \mu_{ref}=\mu_{ref}'-\mu_{ref}$, the reaction (\ref{eq:reac}) will advance or retroced, modifying the concentrations of $J$ by negligible quantities but creating a macroscopic variation of $\phi_{sol}$, the electrostatic potential of the solution. If we call $\Delta \phi_{sol}$ such a variation, elementary thermodynamics gives:
\begin{equation}
\mu_{j}'-\mu_j=\Delta \mu_{j}=e c_j \Delta \phi_{sol}
\label{eq:dmu}
\end{equation}
The condition for the new thermodynamics equilibrium is given by $\sum_j \nu_j \mu_{j}' - n \mu_{ref}'=0$. Taking the difference between this equation and eq.(\ref{eq:ther}), one obtains:
\begin{equation}
\sum_j \nu_j \Delta \mu_{j} - n \Delta \mu_{ref} = 0
\label{eq:diff}
\end{equation}
Using the expression of $\Delta \mu_j$ as given in eq.(\ref{eq:dmu}), eq.(\ref{eq:diff}) becomes:
\begin{equation}
e\Delta \phi_{sol}\sum_j \nu_j c_{j} - n \Delta \mu_{ref} = 0
\end{equation}
Exploiting the charge balance eq.(\ref{eq:bal}), we finally get:
\begin{equation}
-e\Delta \phi_{sol}=\Delta \mu_{ref} 
\label{eq:fin}
\end{equation}

\section{Numerical Methods}

In the article, we have briefly described the model used to calculate the electrostatic potential distribution (and thus $\alpha_{tip}$, $\alpha_{sub}$ and $\beta$) for Azurin in the ECSTM setup. Here, technical details on these calculations will be given. We recall that we are using a continuum model to describe the system: the electrodes (tip and substrate) are assumed to be perfect conductors (the substrate is flat, the tip has a spherical curvature), the protein is a complex-shaped dielectric that cannot be penetrated by ions and the ionic solution is a dielectric that fills all the space left empty by the other components of the system. In particular, the presence of an ionic atmosphere is taken into account by using the linearized Poisson-Boltzman equation eq.(2), which reduces to the Poisson one inside the protein. 
Thus, to calculate the electrostatic potential distribution in the system, we have to solve these equations with the boundary conditions given by the applied tip and substrate potentials. The numerical procedure used to perform such a task is the Finite Difference (FD) method. Basically, the differential Poisson-Boltzmann equation is discretized on a regular grid, using standard expressions for the Laplacian (correct to the third order in the grid spacing, for the present article). Such a discretization translates the differential Poisson Boltzmann equation in a linear problem of the kind $\mathbf{Ax}=\mathbf{b}$, where the known term $\mathbf{b}$ is given by the electrostatic potential at the boundary of the considered system, the unknown $\mathbf{x}$ is the potential in the interior of the system and the coefficient matrix $\mathbf{A}$ depends on the dielectric constant $\epsilon$ and the inverse screening length $k$. 
The boundary condition between the protein and the solution can be directly taken into account in the resolution of the problem by defining a position-dependent dielectric constant and inverse screening length in such a way that $\epsilon(\vec{r})=\epsilon_{pro}$ and $k(\vec{r})=0$ if $\vec{r} \in$ protein, and $\epsilon(\vec{r})=\epsilon_{sol}$  $k(\vec{r})=k_{sol}$ if $\vec{r} \notin$ protein. We have defined the protein region as the interior part of the solvent-excluded surface of the protein.
In particular, in all the simulations we have used $\epsilon_{pro}=4$, a typical value for electrostatic calculations in protein media. 
The step change of properties at the protein boundary is incompatible with the FD approach, where quantities should not vary too much on the grid spacing. For this reason, we smeared $\epsilon(\vec{r})$ and $k(\vec{r})$ at the boundaries by interpolating between the protein and the solution values of $\epsilon$ and $k$.  The smearing is obtained via the arctan function.  

To improve the description of the solution regions close to the electrodes (i.e., the Helmholtz layers), for the region outside the protein we have assumed that:
\begin{eqnarray}
\nonumber \epsilon(\vec{r})=6.0 ~~ k(\vec{r})=0 &~~\mathrm{for}~~&  d < 2.8 \mathrm{\AA} \\
\nonumber \epsilon(\vec{r})=30.0 ~~ k(\vec{r})=k_{sol} &~~\mathrm{for}~~& 2.8 \mathrm{\AA} < d < 4.8 \mathrm{\AA} \\
\epsilon(\vec{r})=78.39 ~~ k(\vec{r})=k_{sol}&~~\mathrm{for}~~&  2.8 \mathrm{\AA} < d < 4.8 \mathrm{\AA}
\end{eqnarray}
where $d$ is the distance of the point $\vec{r}$ from the surface of the closest electrode. The distances and the dielectric constants are typical values used to reproduce the experimental capacitance of double layers.\cite{hamann98} They are reminiscences, in the framework of our very simple model, of the properties of the inner ($d \approx 2.8$ {\AA}) and outer ($d \approx 4.8$ {\AA}) Helmholtz planes. 
Again, at the boundaries between the various $d$ regions, $\epsilon$ and $k$ have been smoothly interpolated.

At this point, we would like to discuss which boundary conditions we used for the electrostatic potential. We have discretized a rectangular box having the substrate and the tip surfaces as basis, and containing the protein at the center. On tip and substrate surfaces, we imposed a value of the $\phi_{sub/tip}$ potential given by $\phi_{sub/tip}=E_{sub/tip}-E^{pzc}_{sub/tip}$, taking $\phi_{sol}=0$ in eq.(2). On the lateral faces we imposed the potential distribution proper for the system without the protein. We enlarged the box until the potential results in the protein region did not depend on the box size anymore.

We remark that the potential distribution for any combination of $\phi_{sub}$ and $\phi_{tip}$ values can be obtained with just two calculations: one with $\phi_{sub}=1$  and $\phi_{tip}=0$ and the other with $\phi_{sub}=0$ and $\phi_{tip}=0$. Since the electrostatic problem that we solved is linear, the potential distribution for any other value of $\phi_{sub}$ and $\phi_{tip}$ can be obtained by linearly combining the results of these calculations. 

Finally, we give some details on the numerical resolution of the non-linear PB equation, necessary to obtain the results presented in Sect.\ref{sec:nonl}. An iterative procedure starting from the linearized solution were employed. At each step, the non-linear terms of the PB equation were Taylor-expanded to the first order in the potential, using the potential obtained in the previous step as the center of the expansion. The so-obtained linear equation is numerically solved and a new iteration is started. The procedure was stopped when the potential values changed less than a given relative threshold ($~10^{-5}$).     
To test the correctness of the results, we checked that the potential profiles presented in Ref.\citenum{kornyshev06} were reproduced by our code. 

We conclude this section by mentioning that the model described here has been implemented in an home-made FORTRAN90 code, and that the solution of the numerical FD problem  has been obtained by standard numerical techniques (bi-conjugate gradient algorithm). Typical parameters used in the calculations are: a grid space of $0.65$ {\AA}, $\approx$ 300 grid points in the $x$ and $y$ directions (parallel to the electrode surface) and 60-80 grid points in the $z$ direction (perpendicular to the electrode surface).


\begin{thebibliography}{99}
\bibitem{foot1} We remark that experimental ECSTM setups usually have a counter electrode in addition to the reference one. The main aim of the counter electrode is to avoid currents to flow in the reference electrode during measurements, since this can alter the reference electrode potential. We shall not deal with such a complication here, since our discussion is focused on electrostatic effects that are not related to current flowing.
\bibitem{foot2} Note that the quantity of released/captured ions needed to recover the equilibrium is very small, negligible w.r.t the ionic concentration already present in solution. Thus, the change in the electrochemical potential directly induced by the variation in the species concentrations, notwithstanding electrostatic effects, is very small (non-polarizable electrode). To give an estimate of concentration changes, let us consider a spherical container for the ionic solution, having radius $R=10$ cm. If the applied voltage changes by 1 V, the charge $Q$ that should be released in solution is $4\pi\epsilon_0 R$. In the specific example, $Q=11$ pC, which means 0.11 fmol of a monovalent ion and a nominal concentration variation of 26 aM!
\bibitem{foot3} For the sake of completeness, we should add that the electrode-solution potential drop also has other components:\cite{bockris00} one is due to the ordered dipole layer formed by the water molecule on the electrode surface, the other is related to the electronic structure of the electrode surface itself.\cite{schmickler96} However, unless the redox center is really close to the electrode (2-3 {\AA}) these potential drops do not significantly contribute to the difference between the electrostatic potential acting on the molecule and in the solution bulk.
\bibitem{foot4} Actually, we have found that when reasonable values for the tip curvature radius are used, the results are not very different from those with a planar tip. Thus, all the following calculations have been done with a planar tip (i.e., infinite curvature radius).
\end{thebibliography}
\end{document}